\documentclass[cameraready]{Interspeech}

\title{WhisperVC: Decoupled Cross-Domain Alignment and Speech Generation for Low-Resource Whisper-to-Normal Conversion}

\author[affiliation={1,2}, orcid=0000-0000-0000-0000]{Dong}{Liu}
\author[affiliation={3,1}]{Juan}{Liu}
\author[affiliation={4}]{Wei}{Ju}
\author[affiliation={4}]{Yao}{Tian}
\author[affiliation={2,3}, orcid=0000-0000-0000-1111, correspondingauthor]{Ming}{Li}

\address{
    $^1$ School of Computer Science, Wuhan University, Wuhan, China \\
    $^2$ School of Artificial Intelligence, The Chinese University of Hong Kong, Shenzhen, China \\
    $^3$ School of Artificial Intelligence, Wuhan University, Wuhan, China \\
    $^4$ AI Center, OPPO, Beijing, China
}

\email{dong.liu@whu.edu.cn}


\usepackage{comment}

\usepackage{amsmath}

\usepackage{graphicx}
\usepackage{booktabs}
\usepackage{multirow}

\begin{document}

\maketitle

\begin{abstract}

Whispered speech lacks vocal-fold excitation, making the intelligible conversion challenging. We propose \emph{WhisperVC}, a three-stage framework for low-resource Whisper-to-Normal (W2N) conversion that decouples cross-domain alignment from speech generation. Stage 1 uses limited paired whisper–normal data with a content encoder and a Conformer-based variational autoencoder (VAE) with soft-DTW alignment to learn domain-invariant semantic representations. Stage 2, trained only on normal speech, employs a Length–Channel Aligner and a two-stage speaker-conditioned mel generator for timbre and prosody modeling. Stage 3 fine-tunes a HiFi-GAN vocoder for waveform synthesis. Experimental results on AISHELL6-Whisper show competitive quality (\textbf{DNSMOS 3.07}, \textbf{UTMOS 2.83}, \textbf{CER 16.93\%}) and WavLM speaker similarity (0.95). The framework also supports privacy-preserving communication as well as non-vocal communication and a rehabilitation tool for post-surgical vocal-fold patients. Samples are available online\footnote{Demo url: https://demo-whispervc.github.io/demo-whispervc/.}.

\end{abstract}

\begin{figure*}
    \centering
    \includegraphics[width=0.75\textwidth]{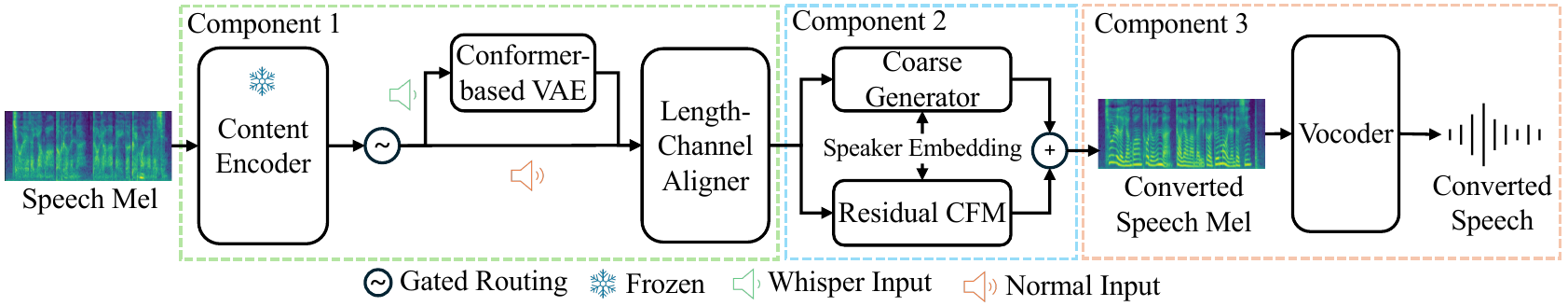}
    \vspace{-2mm}
    \caption{Overview of the proposed whisper-to-normal voice conversion framework.
    }
    \label{fig:framework}
    \vspace{-4mm}
\end{figure*}

\begin{figure*}[t]
    \centering
    \includegraphics[width=0.65\textwidth]{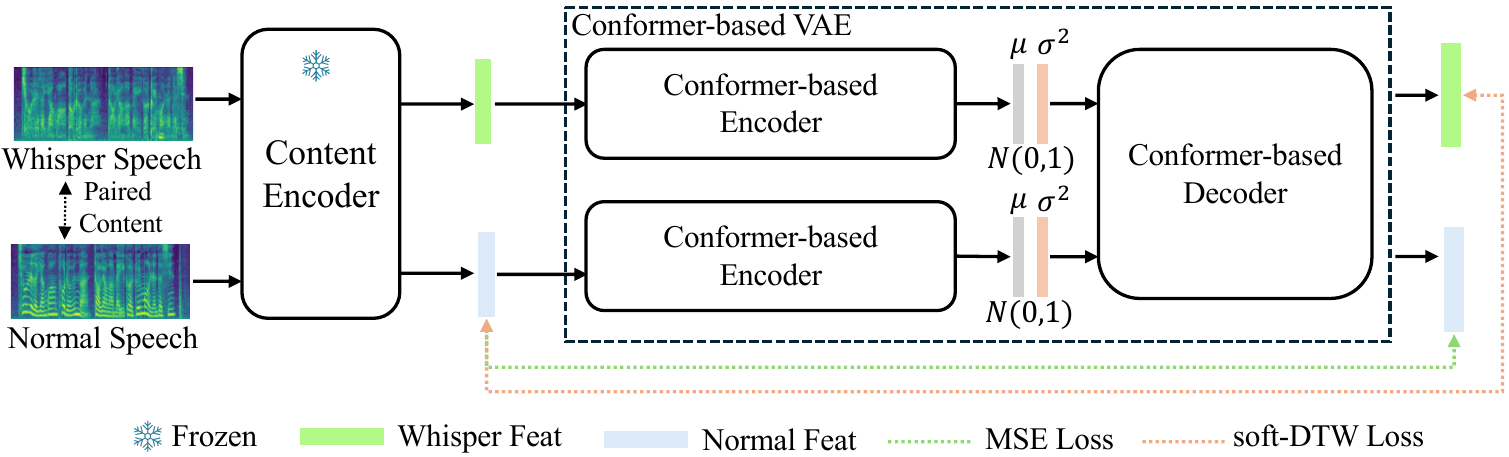}
    \vspace{-2mm}
    \caption{Overview of the proposed Conformer-based VAE module.}
    \label{fig:VAE_framework}
    \vspace{-4mm}
\end{figure*}

\vspace{-1mm}
\section{Introduction}

Whispered speech lacks vocal-fold excitation and exhibits reduced energy and shifted formant frequencies, leading to severe degradation in intelligibility and naturalness. Converting whispered speech into normal voiced speech—referred to as \emph{whisper-to-normal} (W2N) conversion—can benefit individuals with voice disorders or users who must speak quietly in noise-sensitive environments.

W2N remains challenging due to the absence of F0, large spectral mismatch between whisper and normal speech, and temporal inconsistencies across speaking styles. Moreover, parallel whisper–voiced speech corpora are scarce, motivating non-parallel and robust modeling strategies.

Existing approaches are predominantly data-driven. Early neural models employ adversarial or sequence-to-sequence architectures to directly map whispered acoustic features to normal-speech representations. For example, attention-guided GANs~\cite{gao2023novel} and modified Transformer networks~\cite{niranjan2020end} predict mel-spectrograms that are subsequently synthesized by neural vocoders, while comparative studies further demonstrate the effectiveness of GAN-based frameworks~\cite{wagner2024generative}. To alleviate the reliance on strictly parallel data, non-parallel methods based on auxiliary classifier VAEs~\cite{seki2023non} and cycle-consistent GANs~\cite{joysingh2025maskcyclegan} have been proposed, treating W2N as a cross-domain style transfer problem. More recently, lightweight and zero-/one-shot pipelines leveraging self-supervised speech representations have been explored, such as WESPER~\cite{rekimoto2023wesper} and DistillW2N~\cite{tan2025distillw2n}. In contrast, model-driven approaches explicitly exploit interpretable speech production mechanisms, including MFCC inversion~\cite{zhu2022whispered} and source--filter-based excitation reconstruction~\cite{perrotin2020glottal}. While these methods provide interpretability, accurately restoring natural voicing and prosody remains difficult.

Most existing W2N systems adopt a single-stage acoustic mapping framework that jointly learns whisper--normal alignment, speaker conditioning, and acoustic generation. However, the large spectral and temporal mismatch between whisper and normal speech makes stable voicing reconstruction difficult with limited training data and often lack robustness when extended to broader voice conversion (VC) tasks.

To address these limitations and inspired by \cite{yang2023electrolaryngeal}, we propose \emph{WhisperVC}, a decoupled coarse-to-fine framework for W2N conversion. 
Moreover, WhisperVC separates cross-domain alignment from normal-speech generation, enabling unified modeling of W2N and conventional VC within a single architecture. 
Our contributions are threefold:

\begin{itemize}
    \item \textbf{Whisper-specific domain alignment.}
    We introduce a continuous dual-encoder VAE with soft-DTW regularization built upon content encoder representations to model cross-domain alignment between whispered and normal speech, providing stable inputs for downstream generation.

    \item \textbf{Decoupled coarse-to-fine residual generation.}
    We adopt a two-stage generation strategy where a deterministic decoder first predicts a coarse mel representation, followed by an optimal-transport conditional flow matching (OT-CFM) module that models the residual between the coarse prediction and the ground-truth mel, enabling coarse-to-fine refinement of the acoustic representation. 
    A gated dual-path routing mechanism further enables whispered inputs to undergo domain alignment while allowing normal inputs to bypass this stage, unifying W2N and conventional VC within a single framework.

    \item \textbf{Vocoder adaptation for distribution consistency.}
    We fine-tune HiFi-GAN on generated mel-spectrograms to reduce the train--test distribution mismatch between predicted and real acoustic features.
\end{itemize}

By combining explicit cross-domain alignment, gated dual-path routing, and residual flow-based refinement, WhisperVC provides a unified framework for whisper-to-normal conversion while preserving normal-speech voice conversion capability.

\vspace{-2mm}
\section{Methods}

\vspace{-1mm}
\subsection{Framework Overview}

WhisperVC consists of three sequential components:
(1) a whisper-specific domain alignment module,
(2) a coarse-to-fine mel generation framework operating in the normal-speech space,
and (3) a neural vocoder for waveform synthesis.

Given an input utterance $x$ (whispered or normal), our goal is to generate a natural 22.05~kHz waveform $\hat{y}$ while preserving linguistic content and speaker timbre.
The final mel-spectrogram is formulated as

\begin{equation}
\hat{M} = M_c + \hat{R},
\end{equation}

where $M_c$ denotes a coarse mel representation capturing global acoustic structure, and $\hat{R}$ is the residual predicted by the refinement module to model fine-grained acoustic details.

During inference, whispered inputs first undergo domain alignment before mel generation,
while normal inputs bypass the alignment stage through a gated routing mechanism.

\vspace{-1mm}
\subsection{Whisper-Specific Domain Alignment}

We adopt a pretrained content encoder (Whisper-large V3), fine-tuned on a Mandarin whispered--normal corpus~\cite{li2025aishell6}, 
to extract 1280-d content representations $C$ from 16~kHz audio.

Due to representation and temporal mismatch between whispered and normal content features,
we introduce a Conformer-based continuous variational autoencoder (VAE)
to model cross-domain alignment.

The VAE consists of dual encoders and a shared decoder.
Given paired content features $C_w$ (whisper) and $C_n$ (normal),
the encoders produce latent posteriors:

\begin{equation}
q_w(z|C_w)=E_w(C_w), \quad q_n(z|C_n)=E_n(C_n).
\end{equation}

Latent samples $z_w$ and $z_n$ are then decoded to reconstruct aligned features:

\begin{equation}
\hat{C}_w = D(z_w), \quad \hat{C}_n = D(z_n).
\end{equation}

The training objective is

\begin{align}
\mathcal{L}_{\text{VAE}}
&=
\lambda_{\text{KL}}
\left[
\mathrm{KL}(q_w || \mathcal{N}(0,I))
+
\mathrm{KL}(q_n || \mathcal{N}(0,I))
\right]
\nonumber\\
&\quad +
\lambda_{\text{rec}}
\lVert \hat{C}_n - C_n \rVert_2^2
+
\lambda_{\text{DTW}}
\mathrm{softDTW}(\hat{C}_w, C_n).
\end{align}

The soft-DTW~\cite{cuturi2017soft} loss aligns reconstructed whisper features
with normal features under temporal flexibility,
encouraging alignment toward the normal-speech space.

\vspace{-1mm}
\subsection{Coarse-to-Fine Residual Generation}

\textbf{Length--Channel Alignment (LCA).}\\
The content encoder operates on 16~kHz audio,
while mel-spectrograms used for HiFi-GAN synthesis
are extracted at 22.05~kHz.
This results in a length mismatch between encoder features
and mel frames.
To bridge this gap without introducing explicit duration modeling,
we linearly interpolate the encoder features to match the mel length.

Given encoder features $C \in \mathbb{R}^{T_{\text{enc}} \times d}$,
we obtain length-aligned features $\tilde{C}$
via interpolation, followed by a convolutional projection
to map features into the acoustic decoder dimension.

\vspace{1mm}\noindent\textbf{Coarse Mel Generation.}\\
A feed-forward Transformer-based acoustic decoder
predicts a deterministic coarse mel-spectrogram:

\begin{equation}
M_c = G_{\text{coarse}}(\tilde{C}, s),
\end{equation}

where $s$ denotes a 256-dimensional speaker embedding extracted using a SimAM--ResNet34 encoder~\cite{wang2023wespeaker}
pretrained on VoxBlink2~\cite{lin2024voxblink2} and fine-tuned on VoxCeleb2~\cite{chung2018voxceleb2}.

The coarse generator is trained with an $\ell_1$ reconstruction loss:

\begin{equation}
\mathcal{L}_{\text{coarse}} = \lVert M_c - M \rVert_1.
\end{equation}

\vspace{1mm}\noindent\textbf{Residual OT-CFM Refinement.}\\
To improve acoustic fidelity, we model the residual between
the ground-truth mel-spectrogram $M$ and the coarse prediction $M_c$:

\begin{equation}
R = M - M_c .
\end{equation}

Instead of directly generating the full mel-spectrogram,
the residual distribution is modeled using
optimal-transport conditional flow matching (OT-CFM).
Specifically, Gaussian noise $z \sim \mathcal{N}(0,I)$ is transported
to the residual $R$ along a linear interpolation path

\begin{equation}
y_t = (1 - t)z + tR,
\quad t \sim \mathcal{U}(0,1).
\end{equation}

The flow network $f_\theta$ predicts the velocity field
conditioned on the interpolated state $y_t$,
the time step $t$, the length-aligned content feature
$\tilde{C}$, and the speaker embedding $s$.
Training minimizes the velocity matching objective

\begin{equation}
\mathcal{L}_{\text{CFM}}
=
\mathbb{E}_{R,z,t}
\left[
\left\lVert
f_\theta(y_t, t, \tilde{C}, s)
-
(R - z)
\right\rVert_2^2
\right].
\end{equation}



This coarse-to-fine formulation separates global structure modeling
from stochastic detail refinement,
improving stability under cross-domain mismatch.

\textbf{Gated Dual-Path Routing}

To enable unified modeling of W2N and conventional VC,
we introduce a lightweight sigmoid classifier
that predicts a domain indicator and determines whether
the VAE alignment module is applied.

Given content features $C$,
the routing module produces aligned features

\begin{equation}
\tilde{C} =
\begin{cases}
D(E_w(C)), & \text{if classified as whisper}, \\
C, & \text{otherwise}.
\end{cases}
\label{euqa_11}
\end{equation}

The classifier is trained using paired whisper-normal supervision.
This selective alignment corrects cross-domain mismatch
while preserving normal-speech representations.

\vspace{-1mm}
\subsection{Vocoder Adaptation}

We adopt HiFi-GAN~\cite{kong2020hifi} as the neural vocoder.
To reduce train–test mismatch introduced by residual refinement,
the vocoder is fine-tuned on predicted mel-spectrograms.

\begin{table*}[t]
\centering
\caption{W2N performance and ablation results on AISHELL6-Whisper testing set. Best results are highlighted in bold.}
\vspace{-3mm}
\label{tab:cn_w2n_main}
\setlength{\tabcolsep}{3pt}
\small
\resizebox{\textwidth}{!}{
\begin{tabular}{l|ccccccc|c|ccc|c}
\toprule
& \multicolumn{7}{c|}{Naturalness / Quality $\uparrow$}
& \multicolumn{1}{c|}{Intelligibility $\downarrow$}
& \multicolumn{3}{c|}{Speaker Similarity $\uparrow$}
& \multicolumn{1}{c}{Semantic $\uparrow$} \\

Model
& DNSMOS$_{ovrl}$ 
& DNSMOS$_{sig}$ 
& DNSMOS$_{bak}$ 
& P808 
& UTMOS 
& WVMOS 
& NISQA
& CER
& SECS
& WeSpeaker
& WavLM
& SBERT \\

\midrule
Whispered input
& 1.102 & 1.155 & 1.195 & 2.703 & 1.308 & -0.067 & 1.900
& 22.937
& 0.582 & 0.498 & 0.784
& 0.673 \\

Seed-VC~\cite{liu2024zero} (zero-shot)
& 2.868 & 3.255 & 4.031 & 3.740 & 2.467 & 2.814 & 3.088
& 46.423
& 0.851 & 0.772 & 0.952
& 0.758 \\

\midrule
\multicolumn{13}{l}{\textit{Ablation studies}} \\

\quad Coarse-only
& 2.716 & 3.338 & 3.343 & 3.258 & 2.797 & 3.186 & 2.418
& 18.729
& 0.529 & 0.638 & 0.943
& 0.761 \\

\quad \quad + OT-CFM (Full)
& 2.548 & 3.178 & 3.188 & 3.398 & 2.170 & 2.716 & 2.935
& 19.576
& 0.515 & 0.613 & 0.930
& 0.744 \\

\quad \quad + OT-CFM (Residual)
& 2.652 & 3.253 & 3.313 & 3.425 & 2.795 & 3.139 & 2.830
& 18.266
& 0.528 & 0.650 & 0.944
& 0.760 \\

\quad OT-CFM (Residual) w/o VAE
& 1.830 & 3.218 & 1.632 & 2.755 & 1.729 & 1.519 & 2.233
& 40.155
& 0.511 & 0.557 & 0.898
& 0.712 \\

\midrule
\textbf{WhisperVC (Proposed)}
& \textbf{3.072} & \textbf{3.506} & 3.720 & 3.562 & \textbf{2.831} & \textbf{3.352} & 3.164
& \textbf{16.932}
& 0.816 & 0.675 & 0.945
& \textbf{0.785} \\

\midrule
Ground Truth
& 3.141 & 3.552 & 3.795 & 3.680 & 2.868 & 3.129 & 3.395
& -
& 1.000 & 1.000 & 1.000
& 1.000 \\

\bottomrule
\end{tabular}
}
\vspace{-5mm}
\end{table*}

\vspace{-1mm}
\section{Experimental Results}

\vspace{-1mm}
\subsection{Experimental Setup}

WhisperVC consists of three components:
(1) Whisper-Specific Domain Alignment,
(2) Coarse-to-Fine Residual Generation, and
(3) Vocoder Adaptation.
The components are trained sequentially and jointly used during inference.

\textbf{Mandarin (Primary Evaluation).}
All three components are trained on the AISHELL6-Whisper corpus~\cite{li2025aishell6}, which contains paired whispered-normal Mandarin speech with approximately 30 hours.

Component~1 learns cross-domain alignment between whispered and normal speech features.
Component~2 performs normal-speech acoustic modeling using a coarse mel predictor with OT-CFM-based residual refinement.
Component~3 fine-tunes HiFi-GAN to adapt waveform synthesis to the reconstructed mel distribution.

W2N performance is evaluated on whispered test utterances\footnote{All training and testing files are released on the demo page.} to measure reconstruction quality.

To further verify that the gated architecture preserves speaker-conditioned generation capability, we additionally evaluate normal speech VC on AISHELL6-Whisper.
In this setting, normal speech serves as input while a target speaker embedding is provided, examining whether the unified framework maintains standard VC functionality alongside W2N conversion.

Since no prior W2N systems have been reported on AISHELL6-Whisper, we include Seed-VC as a representative generic VC baseline.
For W2N evaluation, whispered speech is directly used as input to Seed-VC, reflecting whisper--normal mismatch.
For VC evaluation, normal speech is used following the standard VC protocol.
Seed-VC is not specifically trained for whisper-to-normal conversion.

\textbf{English (Additional Evaluation).}
The three components are trained on separate corpora to evaluate the effectiveness of the decoupled training strategy.

Component~1 is trained on the paired whispered--normal corpus wTIMIT~\cite{lim2011computational}.
Component~2 is trained on LibriTTS-clean~\cite{zen2019libritts}, which contains only normal speech, following the normal-speech-only generation paradigm.
Component~3 is optimized using LibriTTS-clean together with wTIMIT normal speech to improve robustness to domain variation.

To ensure rigorous zero-shot evaluation, wTIMIT is strictly partitioned at the speaker level into disjoint training, validation, and test sets\footnote{File lists are available at the demo page.}.
Validation and test speakers are completely unseen during training, forming an unseen-speaker evaluation protocol that prevents speaker leakage.
All data are organized as paired normal--whisper utterances to maintain consistent supervision during alignment training and evaluation.

Evaluation is conducted on the wTIMIT whispered test set.
We compare with whisper-oriented systems (WESPER~\cite{rekimoto2023wesper}, DistillW2N~\cite{tan2025distillw2n})
and representative generic VC models (Seed-VC~\cite{liu2024zero}, FreeVC~\cite{li2023freevc}) applied in a zero-shot manner to illustrate the limitation of generic VC under the W2N task.

The content encoder operates at 16~kHz, while acoustic modeling and waveform synthesis are performed at 22.05~kHz.

\vspace{-1mm}
\subsection{Evaluation Metrics}

We evaluate WhisperVC from four perspectives:

\textbf{Naturalness.}
DNSMOS (ovrl/sig/bak/p808)~\cite{reddy2021dnsmos}, UTMOS~\cite{saeki2022utmos}, WVMOS~\cite{andreev2023hifi++}, and NISQA~\cite{mittag2021nisqa}
estimate perceptual quality and signal fidelity, providing a comprehensive approximation of human listening experience.

\textbf{Intelligibility.}
Character Error Rate (CER) is computed using OpenAI Whisper-largeV3-turbo~\cite{radford2023robust}.
These metrics measure phonetic reconstruction accuracy and content preservation.

\textbf{Speaker similarity.}
We report SECS (Resemblyzer cosine similarity) and WavLM~\cite{chen2022wavlm} cosine similarity.

\textbf{Semantic consistency.}
SpeechBERTScore~\cite{saeki2024speechbertscore} measures high-level semantic alignment between converted and reference speech.

\vspace{-1mm}
\subsection{Mandarin W2N Results and Ablation Analysis}

To evaluate W2N conversion performance, we conduct experiments on AISHELL6-Whisper, with results reported in Table~\ref{tab:cn_w2n_main}.

Compared with whispered input, WhisperVC substantially improves perceptual quality and intelligibility
(DNSMOS$_{\text{ovrl}}$: 1.102 $\rightarrow$ 3.072; CER: 22.937\% $\rightarrow$ 16.932\%),
demonstrating that the proposed framework effectively reconstructs natural voiced speech from whispered inputs.

Directly applying a general voice conversion model (Seed-VC) to whispered speech results in severe intelligibility degradation (CER 46.423\%),
indicating that generic VC systems cannot properly handle the large acoustic mismatch between whispered and normal speech.

\textbf{Effect of Component 2 (Coarse-to-Fine Residual Generation).}
Using only the deterministic coarse mel generator (Coarse-only) already improves intelligibility (CER 18.729\%),
showing that the acoustic generator can recover normal-speech structure from the aligned representations produced by Component~1.
However, perceptual quality remains limited due to the deterministic prediction.

Introducing OT-CFM refinement further improves mel reconstruction.
The residual formulation outperforms full-mel modeling, indicating that learning structured residual corrections on top of the coarse mel prediction is more stable than directly modeling full mel trajectories.

\textbf{Effect of Component 1 (Whisper-Specific Domain Alignment).}
Removing the VAE alignment module (OT-CFM Residual w/o VAE) leads to drastic performance degradation (CER 40.155\% and large quality drops across MOS predictors).
This confirms that our VAE-based cross-domain alignment between whispered and normal speech representations is essential for reliable W2N generation.

\textbf{Effect of Component 3 (Vocoder Adaptation).}
Fine-tuning HiFi-GAN on predicted mel-spectrograms further improves perceptual quality and intelligibility
(DNSMOS$_{\text{ovrl}}$: 2.652 $\rightarrow$ 3.072; CER: 18.266\% $\rightarrow$ 16.932\%),
showing that reducing the mismatch between predicted mel distributions and vocoder training data improves waveform synthesis quality.

These results demonstrate that the three components jointly improve W2N performance:
VAE enables cross-domain alignment,
Component~2 provides coarse-to-fine acoustic reconstruction,
and finetuning HiFi-GAN improves waveform synthesis robustness.

\begin{table}[t]
\centering
\caption{VC results on AISHELL6-Whisper testing set. Best results are highlighted in bold.}
\vspace{-3mm}
\label{tab:cn_vc}
\setlength{\tabcolsep}{4pt}
\small
\resizebox{\columnwidth}{!}{
\begin{tabular}{l|cc|c|cc}
\toprule
& \multicolumn{2}{c|}{Naturalness $\uparrow$}
& \multicolumn{1}{c|}{Content $\downarrow$}
& \multicolumn{2}{c}{Speaker $\uparrow$} \\
Model
& DNSMOS$_{OVRL}$ & UTMOS
& CER
& SECS & WavLM \\
\midrule
Seed-VC~\cite{liu2024zero} 
& 3.033 & 2.755
& 4.392
& \textbf{0.894} & 0.727 \\

\textbf{WhisperVC}
& \textbf{3.092} & 2.850
& \textbf{3.331}
& 0.778 & \textbf{0.743} \\

\quad w/o Gate (Equa ~\ref{euqa_11}) 
& 3.078 & \textbf{2.859}
& 4.333
& 0.793 & 0.717 \\

\bottomrule
\end{tabular}}
\vspace{-3mm}
\end{table}

\vspace{-1mm}
\subsection{Voice Conversion Capability (Mandarin)}

To verify that the proposed framework maintains standard voice conversion capability, 
we evaluate normal-to-normal VC on AISHELL6-Whisper, 
with results reported in Table~\ref{tab:cn_vc}.

Compared with the strong baseline Seed-VC, WhisperVC achieves comparable perceptual quality while improving content preservation, reducing CER from 4.392 to 3.331.

We further analyze the effect of the gated architecture by removing the Gated Dual-Path Routing mechanism.
Although perceptual quality remains similar, content preservation degrades noticeably (CER 3.331\% $\rightarrow$ 4.333\%), indicating that the gate helps stabilize generation by adaptively routing information between whisper-oriented and normal-speech pathways.

These results demonstrate that WhisperVC preserves voice conversion capability while simultaneously supporting whisper-to-normal reconstruction within a unified framework.

\begin{table}[t]
\centering
\caption{W2N performance on the wTIMIT testing dataset. Best results are highlighted in bold.}
\vspace{-3mm}
\label{tab:en_w2n}
\setlength{\tabcolsep}{4pt}
\small
\resizebox{\columnwidth}{!}{
\begin{tabular}{l|cc|c|cc}
\toprule
& \multicolumn{2}{c|}{Naturalness $\uparrow$}
& Content $\downarrow$
& \multicolumn{2}{c}{Speaker $\uparrow$} \\
Model
& DNSMOS$_{ovrl}$ & UTMOS
& CER
& SECS & WavLM \\
\midrule

FreeVC~\cite{li2023freevc}
& 2.657 & 2.643
& 26.534
& 0.702 & 0.885 \\

Seed-VC~\cite{liu2024zero}
& 3.108 & 3.321
& 16.709
& \textbf{0.839} & \textbf{0.926} \\

\midrule

WESPER~\cite{rekimoto2023wesper}$\dagger$
& \textbf{3.202} & \textbf{3.496}
& 30.724
& 0.531 & 0.543 \\

DistillW2N~\cite{tan2025distillw2n}$\dagger$
& 3.012 & 1.894
& 36.028
& 0.642 & 0.806 \\

\textbf{WhisperVC}
& 2.894 & 3.276
& \textbf{11.389}
& 0.594 & 0.724 \\

\bottomrule
\end{tabular}}
\vspace{-5mm}
\end{table}

\footnotetext{$^\dagger$ Results obtained using the official pretrained model from the authors’ GitHub repository ( https://github.com/rkmt/wesper-demo, https://github.com/tan90xx/distillw2n ) on our test set.}

\vspace{-1mm}
\subsection{English W2N Evaluation}

To evaluate the generalization ability of WhisperVC across languages, we train the same framework on the English wTIMIT and LibriTTS-clean datasets and report W2N results in Table~\ref{tab:en_w2n}.

WhisperVC achieves the best intelligibility among all compared systems, with a CER of 11.389\%.
Compared with whisper-oriented baselines (WESPER and DistillW2N), the proposed method substantially reduces content recognition errors, demonstrating effective whispered-to-normal reconstruction on English speech.

Generic voice conversion models (Seed-VC and FreeVC) show limited effectiveness when directly applied to whispered inputs.
Although these systems can synthesize perceptually plausible speech, their intelligibility remains inferior to WhisperVC (CER 16.709\% and 26.534\%), indicating that generic VC systems cannot fully resolve the acoustic mismatch between whispered and normal speech.

These results suggest that the proposed alignment and acoustic generation framework generalizes well across languages when trained on the target-language data.

\vspace{-2mm}
\section{Conclusion}

This paper presented WhisperVC, a unified coarse-to-fine framework for whisper-to-normal (W2N) conversion that addresses whisper-normal domain mismatch through whisper-specific representation alignment, residual acoustic generation, and vocoder adaptation. Experiments on AISHELL6-Whisper demonstrate substantial improvements in perceptual quality and intelligibility over whispered input and generic VC baselines, while additional results on wTIMIT show that the framework generalizes beyond Mandarin. Future work will explore improving model efficiency and enabling real-time whisper-to-normal conversion.

\vspace{-2mm}
\section{Acknowledgement}
This research is funded in part by the National Natural Science Foundation of China (62571223) and Yangtze River Delta Science and Technology Innovation Community Joint Research Project (2024CSJGG01100) and OPPO. Many thanks for the computational resource provided by the Advanced Computing East China Sub-Center.

\vspace{-2mm}
\section{Generative AI Use Disclosure}

Large Language Models (LLMs) were used solely for manuscript polishing (e.g., rephrasing and grammar checks) to improve clarity and readability. The LLMs were not used for ideation, methodology, experimental design, data analysis, or result interpretation. All scientific content was produced and verified by the authors.




\bibliographystyle{IEEEtran}
\bibliography{mybib}

\end{document}